\documentclass[sigconf]{acmart}
\AtBeginDocument{%
  \providecommand\BibTeX{{%
    \normalfont B\kern-0.5em{\scshape i\kern-0.25em b}\kern-0.8em\TeX}}}

\usepackage{multirow}
\usepackage{geometry}
\usepackage{caption}
\usepackage{multirow}
\usepackage{colortbl}
\usepackage{arydshln}
\usepackage[T1]{fontenc}
\usepackage{graphicx}
\usepackage{multirow}
\usepackage{tabularx}
\usepackage{xcolor}
\usepackage{colortbl}
\usepackage{algpseudocode}
\usepackage{algorithm}

\newcommand{\zj}[1]{\textcolor{black}{#1}}

\usepackage{amsmath}

\usepackage{amssymb}
\usepackage{mathtools}
\usepackage{amsthm}
\usepackage{tipa}
\usepackage{bm}
\usepackage{bbold}
\usepackage{dsfont}
\usepackage{algorithm} 
\usepackage{algpseudocode} 
\usepackage{appendix}
\usepackage{titletoc}

\usepackage{bbding}
\usepackage{fontawesome}

\copyrightyear{2025}
\acmYear{2025}
\setcopyright{cc}
\setcctype{by-sa}
\acmConference[SIGIR '25]{Proceedings of the 48th International ACM SIGIR Conference on Research and Development in Information Retrieval}{July 13--18, 2025}{Padua, Italy}
\acmBooktitle{Proceedings of the 48th International ACM SIGIR Conference on Research and Development in Information Retrieval (SIGIR '25), July 13--18, 2025, Padua, Italy}\acmDOI{10.1145/3726302.3729950}
\acmISBN{979-8-4007-1592-1/2025/07}

\begin{document}

\title{Diffusion Augmented Retrieval: A Training-Free Approach to Interactive Text-to-Image Retrieval}



\author{Zijun Long}
\affiliation{%
  \institution{Hunan University}
  \streetaddress{2 Lushan S Rd}
  \city{Changsha}
  \state{Hunan}
  \country{China}}
  \postcode{410012}
\email{longzijun@hnu.edu.cn}

\author{Kangheng Liang}
\affiliation{%
  \institution{University of Glasgow}
  \streetaddress{University Avenue}
  \city{Glasgow}
  \country{United Kingdom}
  \postcode{G12 8QQ}}
\email{2743944l@student.gla.ac.uk}

\author{Gerardo	Aragon Camarasa}
\affiliation{%
  \institution{University of Glasgow}
  \streetaddress{University Avenue}
  \city{Glasgow}
  \country{United Kingdom}
  \postcode{G12 8QQ}}
\email{Gerardo.AragonCamarasa@glasgow.ac.uk}

\author{Richard	Mccreadie}
\affiliation{%
  \institution{University of Glasgow}
  \streetaddress{University Avenue}
  \city{Glasgow}
  \country{United Kingdom}
  \postcode{G12 8QQ}}
\email{Richard.Mccreadie@glasgow.ac.uk}

\author{Paul Henderson}
\affiliation{%
  \institution{University of Glasgow}
  \streetaddress{University Avenue}
  \city{Glasgow}
  \country{United Kingdom}
  \postcode{G12 8QQ}}
\email{paul.henderson@glasgow.ac.uk}


\renewcommand{\shortauthors}{Zijun Long, Kangheng Liang, Gerardo Aragon Camarasa, Richard Mccreadie, \& Paul Henderson}

\begin{abstract}

Interactive Text-to-image retrieval (I-TIR) is an important enabler for a wide range of state-of-the-art services in domains such as e-commerce and education. However, current methods rely on finetuned Multimodal Large Language Models (MLLMs), which are costly to train and update, and exhibit poor generalizability. This latter issue is of particular concern, as: 1) finetuning narrows the pretrained distribution of MLLMs, thereby reducing generalizability; and 2) I-TIR introduces increasing query diversity and complexity. As a result, I-TIR solutions are highly likely to encounter queries and images not well represented in any training dataset. \zj{To address this, we propose leveraging Diffusion Models (DMs) for text-to-image mapping, to avoid finetuning MLLMs while preserving robust performance on complex queries. Specifically, we introduce \textit{Diffusion Augmented Retrieval (DAR)}, a framework that generates multiple intermediate representations via LLM-based dialogue refinements and DMs, producing a richer depiction of the user's information needs. This augmented representation facilitates more accurate identification of semantically and visually related images.}
Extensive experiments on four benchmarks show that for simple queries, DAR achieves results on par with finetuned I-TIR models, yet without incurring their tuning overhead. Moreover, as queries become more complex through additional conversational turns, DAR surpasses finetuned I-TIR models by up to 7.61\% in Hits@10 after ten turns, illustrating its improved generalization for more intricate queries. \\

\end{abstract}

\vspace{-8mm}
\ccsdesc[500]{Information systems~Novelty in information retrieval}
\vspace{-8mm}
\keywords{Interactive Text-to-image Retrieval, Conversational IR, Diffusion Augmented Retrieval.}
\vspace{-8mm}

\maketitle

\section{Introduction}

\begin{figure}
    \centering
    \includegraphics[width=0.9\linewidth]{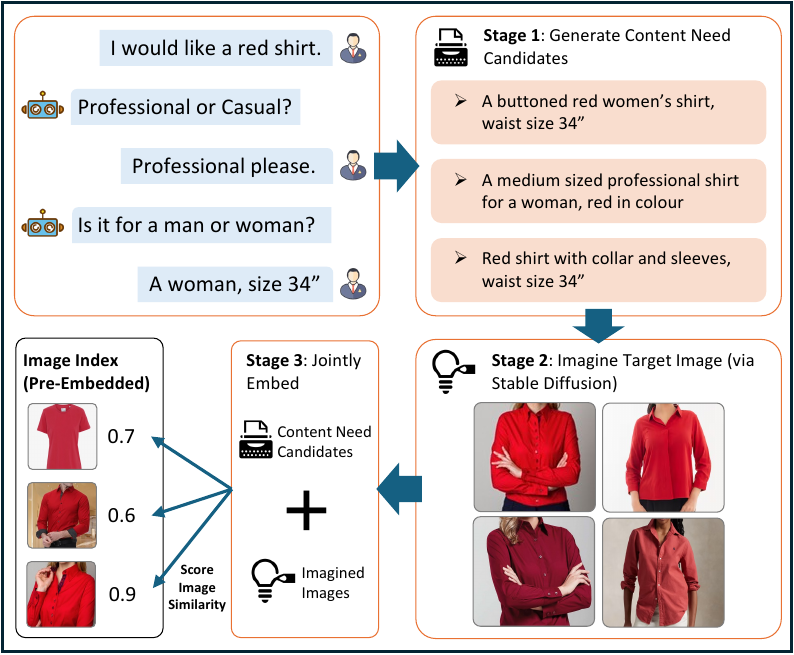}
    \vspace{-2mm}
    \caption{Diffusion  Conceptual Framework}
    \vspace{-2mm}
    \label{fig:DuoGenXSmall}
\end{figure}

\looseness -1 Interactive Text-to-Image Retrieval (I-TIR) seeks to identify relevant images for a user through a turn-by-turn dialogue with a conversational agent. This approach allows users to progressively refine their queries with the agent's guidance, facilitating the retrieval of highly specific results, even when initial queries are vague or simplistic \cite{RN626}, as shown in the top-left of Figure~\ref{fig:DuoGenXSmall}. This conversational paradigm is increasingly popular, as it is well-suited to content exploration use cases, such as fashion shopping or searching for long-tail content, where the agent can help the user refine their search~\cite{RN626,wu2021fashion, saito2023pic2word, wan2024cross,yuan2021conversational,RN155}.

Current state-of-the-art solutions to I-TIR typically rely on finetuning Multimodal Large Language Models (MLLMs) for the retrieval task~\cite{long2024multiway}, aiming to bridge the domain gap between MLLM pretraining and retrieval objectives~\cite{lee2024interactive,RN626,long2024understanding,long2024cfir,long2024clce,ge20243shnet,ge2021structured}. However, we argue that this finetuning is not always needed. First, the one-to-one mapping between text and images that MLLM finetuning aims to establish is neither sufficient nor feasible for complex and diverse queries in I-TIR.
Second, by restricting the pretrained distribution, finetuning compromises the broader generalizability that MLLMs acquire from pretraining, causing such models to underperform for I-TIR whenever they encounter dialogues outside the finetuning distribution. This motivates our key research question: \emph{How can we enhance the generalizability of I-TIR frameworks without additional training?}

To address this research question, we begin by revisiting the motivations for finetuning MLLMs and examining its associated drawbacks. MLLM pretraining typically uses large-scale, noisy text-image pairs from internet or crowdsourced data to learn a joint embedding space. However, this approach often leads to sparse alignment, where text and visual representations of the same content may have differing embeddings. In contrast, retrieval tasks require a more exact, one-to-one mapping to maximize text-image similarity scores. Although finetuning on retrieval datasets partially addresses this gap, we argue that achieving such a high degree of alignment is inherently challenging in I-TIR. Textual queries often lack the fine-grained detail needed to describe target images.  These limitations make it nearly impossible to establish consistent one-to-one mappings between diverse dialogue queries and target images. Indeed, the finetuned mappings remain confined to the training distribution and therefore lack generalizability.

This lack of generalization in finetuned MLLMs poses a particular obstacle to developing robust I-TIR solutions. Multi-turn interactions frequently generate lengthy, varied, and complex dialogues~\cite{zhai2024investigating}, which are difficult for MLLMs to handle effectively when only pretraining or finetuning on limited I-TIR datasets is available. Indeed, I-TIR is more semantically difficult than text-only conversational search because the cross-modal information space is larger, and training data are scarcer~\cite{RN626,murahari2020large}. For instance, with 10 dialogue turns and up to 100 potential questions per turn, the conversation space can reach 100 quintillion distinct trajectories, meaning practically sized training and validation datasets can only cover the scenarios space very sparsely. Consequently, once deployed, finetuned I-TIR approaches are likely to encounter queries and images that are poorly represented in any training set, leading to rapid performance degradation~\cite{Tejero-de-Pablos_2024_WACV}. Hence, we argue that I-TIR approaches should explore how to better leverage zero-shot MLLMs, rather than rely on finetuning them.

In terms of efficiency, finetuning MLLMs at scale requires substantial computational resources and GPU memory, making frequent updates impractical for smaller organizations and research groups~\cite{xu2024survey}. For instance, finetuning MLLMs can demand over 300GB of GPU memory, exceeding the capacity of many standard GPU clusters and effectively rendering additional training infeasible. Consequently, there is a pressing need to explore more efficient strategies to harness the capabilities of advanced MLLMs for I-TIR.


Recent advances in diffusion-based generative models present a promising pathway for adapting pretrained MLLMs to retrieval tasks without finetuning. We argue that DMs provide valuable prior knowledge on the text-to-image mapping—knowledge that pretrained MLLMs do not fully capture and finetuning seeks to acquire. By leveraging DM-generated images, we address the domain gaps between pretraining and retrieval objectives without requiring additional training. Furthermore, generating multiple images as intermediate representations alleviates the challenge of establishing one-to-one text-image mappings, while preserving the cross-modal knowledge of MLLMs to enhance generality.

Building on this insight, we propose a new I-TIR framework, referred to as Diffusion Augmented Retrieval (\emph{DAR}), illustrated in Figure~\ref{fig:DuoGenXSmall}. The core idea underpinning DAR is to produce multiple intermediate representations of the user’s information need, via LLM-based refinement of dialogue~\cite{wu2023brief,touvron2023llama,kumar2024towards} and diffusion-based image generation~\cite{saharia2022photorealistic, Rombach_2022_CVPR}. These generative components collectively imagine the user’s intent based on the conversation, and the images in the target corpus are then ranked according to their similarity to these imagined representations. This multi-faceted portrayal of the query provides a richer, more robust foundation than finetuned MLLMs, leading to more accurate identification of semantically and visually related images. Moreover, DAR is compatible with various LLMs and MLLMs and requires no finetuning (see Section~\ref{sec:compability}), making it well-suited for integration with larger, more powerful MLLMs in the future.

\begin{figure*}[]
    \centering
    \includegraphics[width=1\linewidth]{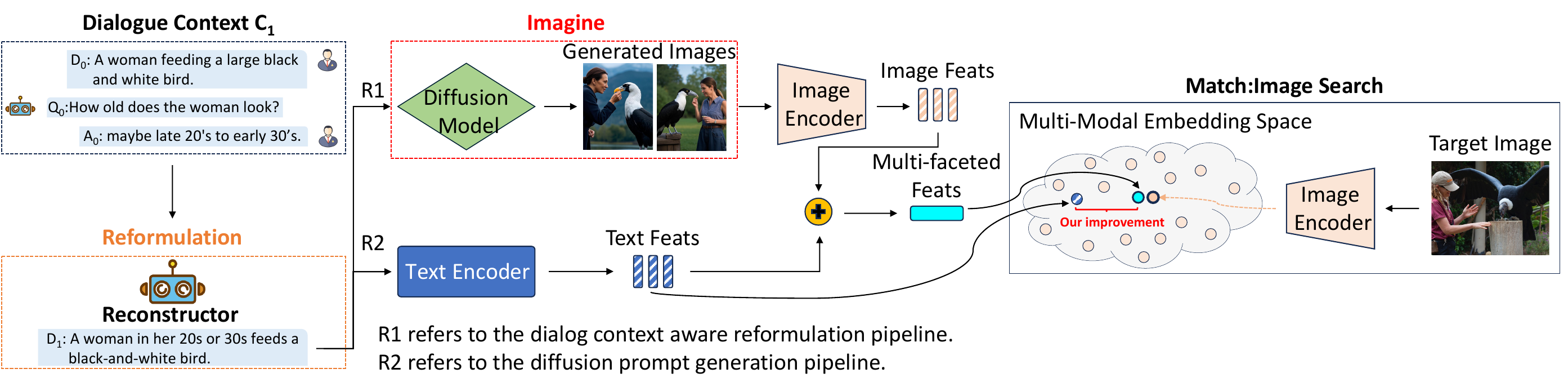}
    \vspace{-4mm}
    \caption{The overall architecture of our proposed framework DAR.}
    \label{fig:overview_archi}
    \vspace{-4mm}
\end{figure*}

In summary, our proposed DAR framework offers the following key contributions:  

\begin{enumerate} \item \textbf{Diffusion Augmented Retrieval (DAR) Framework:}
We introduce a novel I-TIR framework, DAR, which transfers prior text-to-image knowledge from DMs through image generation. This approach bridges the gap between MLLMs’ pretraining tasks and retrieval objectives without requiring finetuning.

\item \textbf{Multi-Faceted Cross-Modal Representations:}  
DAR generates multiple intermediate representations of the user’s information need using LLMs and DMs across interactive turns. By enriching and diversifying how queries are represented, DAR more effectively identifies semantically and visually relevant images.

\item \textbf{Preserving MLLMs’ Generalizability in I-TIR:}  
Our approach maintains the broad cross-modal knowledge captured by MLLMs, enabling strong zero-shot performance. By avoiding finetuning, DAR retains the adaptability and versatility of large pretrained models.


\item \textbf{Comprehensive Empirical Validation:}
We rigorously evaluate DAR on four diverse I-TIR benchmarks, mainly in comparison to prior approaches that rely on finetuned MLLMs. For initial, simpler queries (first conversation turn), DAR achieves performance comparable to state-of-the-art finetuned models. However, as query complexity grows over multiple turns, DAR consistently surpasses finetuned and un-finetuned approaches by up to 4.22\% and 7.61\% in Hits@10, respectively.

\end{enumerate}

\section{Related work}
\subsection{Visual Generative Models}

Before Diffusion Models (DMs), the most successful visual generative models were Generative Adversarial Networks (GANs)~\cite{goodfellow2020generative, goodfellow2014generative, creswell2018generative}, which introduced adversarial training for image synthesis~\cite{goodfellow2020generative}. While GANs achieved notable success in tasks like image generation and style transfer~\cite{goodfellow2020generative}, they suffer from limitations such as mode collapse, training instability, and the need for carefully tuned architectures, making them less robust for diverse tasks~\cite{gui2021review}.

DMs address these challenges by using a noise-adding and noise-removal process, enabling stable and high-quality generation~\cite{croitoru2023diffusion}. Initial approaches like Denoising Diffusion Probabilistic Models laid the groundwork~\cite{ho2020denoising}, followed by advancements such as Stable Diffusion~\cite{Rombach_2022_CVPR} and Imagen~\cite{saharia2022photorealistic}, which enhance efficiency and scalability~\cite{croitoru2023diffusion, saharia2022photorealistic}. Diffusion models excel in generating high-fidelity, diverse outputs without mode collapse and are adaptable to multimodal tasks like text-to-image generation~\cite{ramesh2021zero}. In this work, we aim to leverage the prior knowledge of DMs to bridge the domain gap between the pretraining tasks of Multimodal Large Language Models (MLLMs) and retrieval objectives, achieving superior zero-shot performance and better handling of unseen samples.

\vspace{-2mm}
\subsection{Interactive Text-to-Image Retrieval}

Interactive text-to-image retrieval (I-TIR) overcomes the limitations of conventional single-turn methods by iteratively clarifying a user's information need across multiple dialog turns. This iterative process is especially effective for image retrieval, as a single initial query often fails to capture the fine details of target images. By incorporating user feedback over successive turns, I-TIR progressively aligns with user preferences, improving both retrieval accuracy and user satisfaction. However, this flexibility also adds complexity; the representation of a user’s information need becomes more elaborate due to the diversity of multi-turn dialogs.

Recent studies show the potential of LLMs and MLLMs for I-TIR. For instance, ChatIR~\cite{RN626} employs LLMs to simulate dialogues between users and answer bots, compensating for the scarcity of specialized text-to-image datasets tailored to I-TIR. Although this approach opens a promising direction for research, it does not address the unique training hurdles posed by I-TIR—specifically, handling highly diverse dialogue inputs. PlugIR~\cite{lee2024interactive} further advances I-TIR by improving the diversity of top-\emph{k} results using \(k\)-means clustering to group similar candidate images and identify representative exemplars. However, this technique incurs additional computational overhead. Moreover, ChatIR and PlugIR, as well as other I-TIR strategies, rely on finetuning on small, curated datasets to achieve good results on a specific benchmark~\cite{zhu2024enhancing,RN632,RN633}, limiting their ability to generalize to the broad range of real-world dialogues. Consequently, they often fail on out-of-distribution queries, leading to diminished performance in practical settings, as shown in Section~\ref{sec:robustness}.

To address these challenges, we propose the DAR framework, which prioritizes zero-shot I-TIR performance. By avoiding dataset-specific finetuning altogether, DAR avoids the reduced distribution space introduced by smaller, finetuned datasets, thereby achieving superior generalizability.

\vspace{-2mm}
\section{DAR for Interactive Text-to-Image Retrieval}
\label{sec:method}

In this section, we present the \emph{Diffusion Augmented Retrieval Framework (DAR)}, illustrated in Figure~\ref{fig:overview_archi}, which comprises three main steps: dialogue reformulation, imagination (image generation), and matching.

To bridge the domain gap between the pretraining tasks of multimodal large language models (MLLMs) and Interactive Text-to-Image Retrieval (I-TIR) without additional training, DAR employs two generative models to imagine user intentions and generate intermediate representations for retrieval:
\begin{itemize}
    \item \textbf{Large Language Model (LLM):} An LLM is utilized to adapt the dialogue context, ensuring it closely aligns with the retrieval model’s input requirements and user's intent. This adaptation enhances the relevance of retrieval results by reducing ambiguities in dialogues.
    
    \item \textbf{Diffusion Model (DM):} DMs provide valuable prior knowledge about text-to-image mappings—information that un-finetuned MLLMs lack. By generating multiple images based on LLM refined prompts, DMs create multifaceted representations of the user's intent, thereby bridging the domain gap and eliminating the need for finetuning MLLMs.
\end{itemize}

The integration of these intermediate representations offers a richer and more robust foundation for bridging the text-to-image domain gap than finetuned MLLMs, leading to more accurate identification of semantically and visually related images.

In the following sections, we first provide background on the settings of I-TIR in Section \ref{sec:preliminary}. Next, we discuss the first step of DAR, dialog reformulation in two separate pipelines, in Section \ref{sec:reformulation}, illustrating how they enhance both DM generation and refine dialogues. We then introduce our main contributions in Section \ref{sec:GAR}, focusing on the core concept of diffusion augmented multi-faceted generation. Finally, we describe the detailed retrieval procedure in Section \ref{se:re}.

\vspace{-2mm}
\subsection{Preliminary}
\label{sec:preliminary}
\looseness -1 Interactive text-to-image retrieval is formulated as a multi-turn task that begins with an initial user-provided description, \( D_0 \). The objective is to identify a target image through an iterative dialogue between the user and the retrieval framework. At each turn \( t \), the retrieval framework generates a question \( Q_t \) to clarify the search, and the user responds with an answer \( A_t \). This interaction updates the dialogue context \( C_t = (D_0, Q_1, A_1, \ldots, Q_t, A_t) \), which is processed—such as by concatenating all textual elements—to form a unified search query \( S_t \) for that round. The retrieval framework then matches images \( I \in \mathcal{I} \) in the image database against \( S_t \), ranking them based on a similarity score \( s(I, S_t) \). This process iterates until the target image \( I^* \) is successfully retrieved or the maximum number of turns is reached.  Formally, this process can be defined as:
\[
I^* = \arg\max_{I \in \mathcal{I}} s(I, S_T)
\]
where \( S_T \) is the final search query after \( T \) dialogue turns.

\subsection{Dialog Context Aware Reformulation}
\label{sec:reformulation}

While multi-turn interactions help capture user intent, raw dialogue data can introduce noise and complexity that degrades retrieval performance. Both encoders and diffusion models struggle with lengthy or ambiguous dialogue context, particularly because DMs have limited capacity for long or complex descriptions. Nevertheless, the quality of images generated by DMs is crucial for DAR to achieve superior performance. Moreover, discrepancies between the training distributions of encoders and DMs make it difficult to generate images that accurately reflect user intentions. To address these challenges, we propose two targeted approaches: refining the dialogue for textual representations used in retrieval; and optimizing the prompts for the DM generation process:

\begin{enumerate}
\item \textbf{Dialogue Context Aware Reformulation:}  
This pipeline adapts the dialogue context to better align with the input expectations of encoders and the user's intent. Instead of directly using the raw dialogue context \( C_t = \{ D_0, Q_1, A_1, \ldots, Q_t, A_t \} \) as textual representations, we follow a multi-step process to refine the input. 

\begin{enumerate}
    \item Summarizing the Dialogue: We first ask an LLM to summarize the entire dialogue context \( C_t \), providing a coherent and concise overview of the conversation up to turn \( t \).
    \item Structuring the Input: The summarized dialogue is then reformulated into a specific format that clearly distinguishes the initial query (\( D_0 \)) from the subsequent elaborations in the dialogue. This ensures that the LLM understands the progression of the conversation and the relevance of each turn.
    \item Generating the Refined Query: Using this structured input, we prompt the LLM to generate a refined query \( S_t \) that adheres to the encoders' input distribution. This ensures the generated query captures the user’s intent in a way that facilitates accurate retrieval.
\end{enumerate}

The reformulation process can be expressed as:
\[
S_t = \mathcal{R}_1(C_t)
\]
where \( \mathcal{R}_1 \) denotes the reformulation function utilizing LLMs. An example prompt for \( \mathcal{R}_1 \) is:  "The reconstructed [New Query] should be concise and in an appropriate format to retrieve a target image from a pool of candidate images." This approach allows the dialogue context to be transformed into a more structured and relevant form for the retrieval pipeline, optimizing the alignment between user intent and the model's output.

\item \textbf{Diffusion Prompt Reformulation:}  
This pipeline generates multiple prompts \( P_{t,k} \) for use by the subsequent diffusion models based on the reformulated dialogue \( S_t \). By producing diverse prompts, we ensure that the generated images align with the diffusion model’s training distribution, capturing various linguistic patterns and semantic nuances. The reformulation process follows these steps:

\begin{enumerate}
    \item Structuring the Prompt Template: We begin by structuring a prompt template that captures key elements from the reformulated dialogue \( S_t \), including the primary subject, setting, and important details. This structured format helps guide the diffusion model to generate images that are semantically coherent with the user’s intent.
    
    \item Generating Diverse Prompts: Using the structured template, we generate multiple distinct prompts \( P_{t,k} \) by varying linguistic patterns, modifiers, and details, ensuring a variety of interpretations that reflect the full scope of the dialogue. This diversity helps cover different possible details in the image generation process.
    
    \item Adapting to the DM's Distribution: The generated prompts are further adjusted to align with the diffusion model’s training distribution. This adaptation ensures that the prompts match the model’s expectations and improve the relevance of the generated images.
\end{enumerate}

The reformulation process is expressed as:
\[
P_{t,k} = \mathcal{R}_2(S_t, k) \quad \text{for } k = 1, 2, \ldots, K
\]
where \( \mathcal{R}_2 \) denotes the prompt generation function, and \( K \) is the number of prompts generated per turn. An example template prompt for \( \mathcal{R}_2 \) is: ``[Adjective] [Primary Subject] in [Setting], [Key Details]. Style: photorealistic."

By generating diverse prompts, this approach ensures that the diffusion models produce images that act as multiple
intermediate representations of the user’s information need, which
are both semantically rich and better aligned with the user's query.
\end{enumerate}

Overall, since we focus on zero-shot scenarios where retrieval models are not finetuned on a particular domain or dataset, reformulating the dialog can supply additional context or details. This helps bridge the semantic gap between the user’s language and the system’s learned representation space, boosting performance without further training. Code for these reformulation pipelines is available at \url{https://anonymous.4open.science/r/Diffusion-Augmented-Retrieval-7EF1/README.md}.

\begin{algorithm}
\caption{Overall Retrieval Process of DAR}
\label{alg:duogenx_retrieval}
\begin{algorithmic}[1]
\Require Initial user description \( D_0 \)
\Require Image pool \( \mathcal{I} = \{ I_1, I_2, \ldots, I_N \} \)
\Require Maximum number of turns \( T \)
\Ensure Retrieved target image \( I^* \)

\State Initialize dialogue context \( C_0 = \{ D_0 \} \)

\For{turn \( t = 1 \) to \( T \)}
    \State \textbf{System Inquiry:} Generate question \( Q_t \) based on \( C_{t-1} \)
    
    \State \textbf{User Response:} Receive answer \( A_t \) from the user
    
    \State Update dialogue context: \( C_t = C_{t-1} \cup \{ Q_t, A_t \} \)
    
    \State \textbf{Dialogue Context Aware Reformulation:}
    \State \( S_t \gets \mathcal{R}_1(C_t) \) \Comment{Transform \( C_t \) into caption-style description}
    
    \State \textbf{Diffusion Prompt Reformulation:}
    \For{each prompt index \( k = 1 \) to \( K \)}
        \State Generate prompt \( P_{t,k} \gets \mathcal{R}_2(S_t, k) \)
        \State Generate synthetic image \( \hat{I}_{t,k} \gets \text{DiffusionModel}(P_{t,k}) \)
    \EndFor
    
    \State \textbf{Feature Fusion:}
    \State \( F_t \gets \alpha \cdot E(S_t) + \beta \cdot \left( \sum_{k=1}^{K} E(\hat{I}_{t,k}) \right) \)
    
    \State \textbf{Similarity Computation and Ranking:}
    \For{each image \( I \in \mathcal{I} \)}
        \State Compute similarity score \( s(I, F_t) \gets \text{Similarity}(E(I), F_t) \)
    \EndFor
    \State Rank images in \( \mathcal{I} \) based on \( s(I, F_t) \) in descending order
    
    \State \textbf{Retrieve Top Image:} \( I^*_t \gets \arg\max_{I \in \mathcal{I}} s(I, F_t) \)
    
    \If{ \( I^*_t \) is satisfactory}
        \State \textbf{Terminate Retrieval:} Set \( I^* = I^*_t \)
        \State \textbf{Exit} the loop
    \EndIf
\EndFor

\State \textbf{Final Retrieval:} \( I^* = \arg\max_{I \in \mathcal{I}} s(I, F_T) \)

\end{algorithmic}
\end{algorithm}

\subsection{Diffusion Augmented Multi-Faceted Generation}
\label{sec:GAR}
Following the reformulation step, we obtain a refined textual dialog for retrieval (Section~\ref{sec:reformulation}) and multiple prompts that capture different aspects of the user’s intent. We now proceed to the \emph{imagine} step, where these prompts are used to generate images that augment the retrieval process. We term this approach \emph{Diffusion  Augmented Retrieval (DAR)}.

The core idea of DAR involves utilizing DMs to generate synthetic images that serve as multiple intermediate representations of the user’s information needs, thereby enhancing the retrieval process. DMs excel at producing high-quality, visually realistic images from textual descriptions, enabling them to closely align with the user's intent. By leveraging the prior visual knowledge embedded in DMs through image generation, DAR establishes many-to-one mappings between queries and target images instead of one-to-one mappings. This multi-faceted representation of queries addresses the challenges posed by incomplete or ambiguous textual inputs, offering a richer and more diverse set of visual representations for retrieval. Consequently, DAR achieves robust zero-shot performance, particularly in I-TIR scenarios where labeled data is scarce.

Specifically, for text-guided diffusion generation, the reverse process is conditioned on a text embedding \(\mathbf{t}_d\) from the diffusion text encoder:
\zj{\[
p_\theta(x_{t-1} \mid x_t, \mathbf{t}_d) 
= \mathcal{N}\Bigl(
x_{t-1}; \boldsymbol{\mu}_\theta\bigl(x_t, t, \mathbf{t}_d\bigr),\, 
\Sigma(t)
\Bigr),
\]
}

\zj{where the function \( \boldsymbol{\mu}_\theta(x_t, t, \mathbf{t}_d) \) represents the learned denoiser, which predicts the mean of the posterior distribution for \( x_{t-1} \) given the noisy input \( x_t \), timestep \( t \), and additional context \( \mathbf{t}_d \). } 

Let 
\[
    \bigl\{P_{t,k}\bigr\}_{k=1}^K
\]
be the set of \(K\) diffusion-ready prompts produced by the reformulation process. We feed each prompt \(P_{t,k}\) into the diffusion model \(D(\cdot)\) to generate a corresponding set of images:
\[
    \bigl\{\hat{I}_{t,k}\bigr\}_{k=1}^K 
    = \Bigl\{D\bigl(P_{t,k}\bigr)\Bigr\}_{k=1}^K.
\]
Each generated image \(I_{t,k}\) reflects one possible interpretation of the user’s intent based on the prompt \(P_{t,k}\). By leveraging the large-scale pretraining of the diffusion model and generating multiple images per turn, DAR captures diverse visual features relevant to the query.

\subsection{Retrieval Process}
\label{se:re}

\subsubsection{DAR Encoding and Feature Fusion}
In the DAR framework, an MLLM is employed to encode both the reformulated textual dialogue \(S_t\) and the generated images \(\{I_{t,k}\}_{k=1}^K\). Although MLLMs may be based on unified or two-tower architectures for handling text and images, their exact design does not affect the overall flow of DAR, as we only utilize them as encoders. Consequently, we do not delve into internal implementation details. Instead, we focus on four key steps: \emph{dialog encoding}, \emph{generated image encoding}, \emph{candidate image encoding}, and \emph{feature fusion}.

\begin{enumerate}
    \item \textbf{Dialog Encoding.} 
    Let \(E_{t}(\cdot)\) denote the text encoder of the MLLM for dialogue inputs. Given the reformulated textual dialogue \(S_t\) at turn \(t\), we obtain its embedding as:
    \begin{equation}
        \mathbf{t}_t \;=\; E_{t}(S_t),
    \end{equation}
    where \(\mathbf{t}_t \in \mathbb{R}^{d}\) is the resulting textual embedding and \(d\) is the embedding dimensionality.

    \item \textbf{Generated Image Encoding.} 
    Let \(E_{v}(\cdot)\) denote the image encoder of the MLLM. For each generated image \(\hat{I}_{t,k}\), where \(k = 1, 2, \ldots, K\), its embedding is:
    \begin{equation}
        \mathbf{i}_{t,k} \;=\; E_{v}\bigl(\hat{I}_{t,k}\bigr),
    \end{equation}
    so that \(\{\mathbf{i}_{t,k}\}_{k=1}^K\) represents the set of embeddings corresponding to the \(K\) synthetic images generated at turn \(t\).

    \item \textbf{Image Encoding for Candidate Images.} 
    Let \(\mathcal{I} = \{I_1, I_2, \ldots, I_N\}\) be the set of \(N\) candidate images in the database. We encode each candidate image \(I_j\) using the same image encoder \(E_{v}(\cdot)\) that we used for generated images:
    \begin{equation}
        \mathbf{i}_j \;=\; E_{v}(I_j), \quad j = 1, 2, \ldots, N.
    \end{equation}
    This ensures all images, whether generated or from the database, reside in the same embedding space.

    \item \textbf{Feature Fusion.}
    To create a multi-faceted feature representation \(F_t\) at turn \(t\), we integrate the textual embedding \(\mathbf{t}_t\) the aggregated embeddings of the generated images \(\{\mathbf{i}_{t,k}\}\). We introduce weighting factors \(\alpha\) and \(\beta\) to balance the contributions of the textual and visual embeddings, respectively. Formally:
    \begin{equation}
        F_t \;=\; \alpha\, \mathbf{t}_t
        \;+\;\beta\, \Bigl(\sum_{k=1}^{K} \mathbf{i}_{t,k}\Bigr),
        \quad \alpha + \beta = 1.
    \end{equation}
    Here, \(F_t \in \mathbb{R}^d\) is the fused feature vector that captures both multi-faceted linguistic and visual semantics. By adjusting \(\alpha\) and \(\beta\), one can control the relative influence of textual and visual information in the final representations.
\end{enumerate}

The above steps ensure that DAR leverages the complementary strengths of MLLMs and generated contents, thereby enhancing retrieval accuracy in zero-shot settings without requiring further model training.

\subsubsection{Matching.}
The matching process in DAR leverages the multi-faceted feature representation \( F_t \) to identify the most relevant images from the image pool \(\mathcal{I} = \{ I_1, I_2, \ldots, I_N \}\). This process involves the following steps:

\begin{enumerate}
    \item \textbf{Similarity Computation:} 
    For each candidate image \( i_{j} \in \mathcal{I} \), we compute the cosine similarity score \( s(I, F_t) \) between \(\mathbf{i}\) and the fused feature \( F_t \) as follows:

    \[
    s\bigl(\mathbf{I}, \mathbf{F}_t\bigr) 
    = \frac{\mathbf{i_{j}} \cdot \mathbf{F}_t}{\lVert \mathbf{i_{j}} \rVert \, \lVert \mathbf{F}_t \rVert}.
    \]

    \item \textbf{Ranking:}
    We then rank the images in descending order of their similarity scores and retrieve the top-\(k\) results:

    \begin{align}
    [I^*_1, I^*_2, \dots, I^*_k] &=
    \mathrm{top}^k_{I \in \mathcal{I}}\, s(I, F_t),
    \end{align}

    where \(\mathrm{top}^k(\cdot)\) returns the \(k\) highest-scoring images. By comparing the fused embedding \(F_t\) with each candidate image embedding \(\mathbf{i}\), this step identifies those images that best match both the refined query and the diffusion-generated features.

    \item \textbf{Iteration:}
    Finally, the retrieval process iterates with new dialogue turns, updating \( F_{t} \) at each turn \( t \), until the maximum number of turns \( T \) is reached or the target image \( I^* \) is successfully retrieved. At the final turn \( T \), the retrieval result is given by

    \begin{equation}
        I^* \;= \arg\max_{I \in \mathcal{I}} \, s(I, F_T).
    \end{equation}
    
\end{enumerate}

\zj{The QA turn generation follows the methods outlined in \cite{RN626}, with further details provided in Section \ref{sec:exp_setting}.} The complete retrieval process of DAR is given in Algorithm~\ref{alg:duogenx_retrieval}.

\begin{figure*}[]
    \centering
    \includegraphics[width=0.9\linewidth]{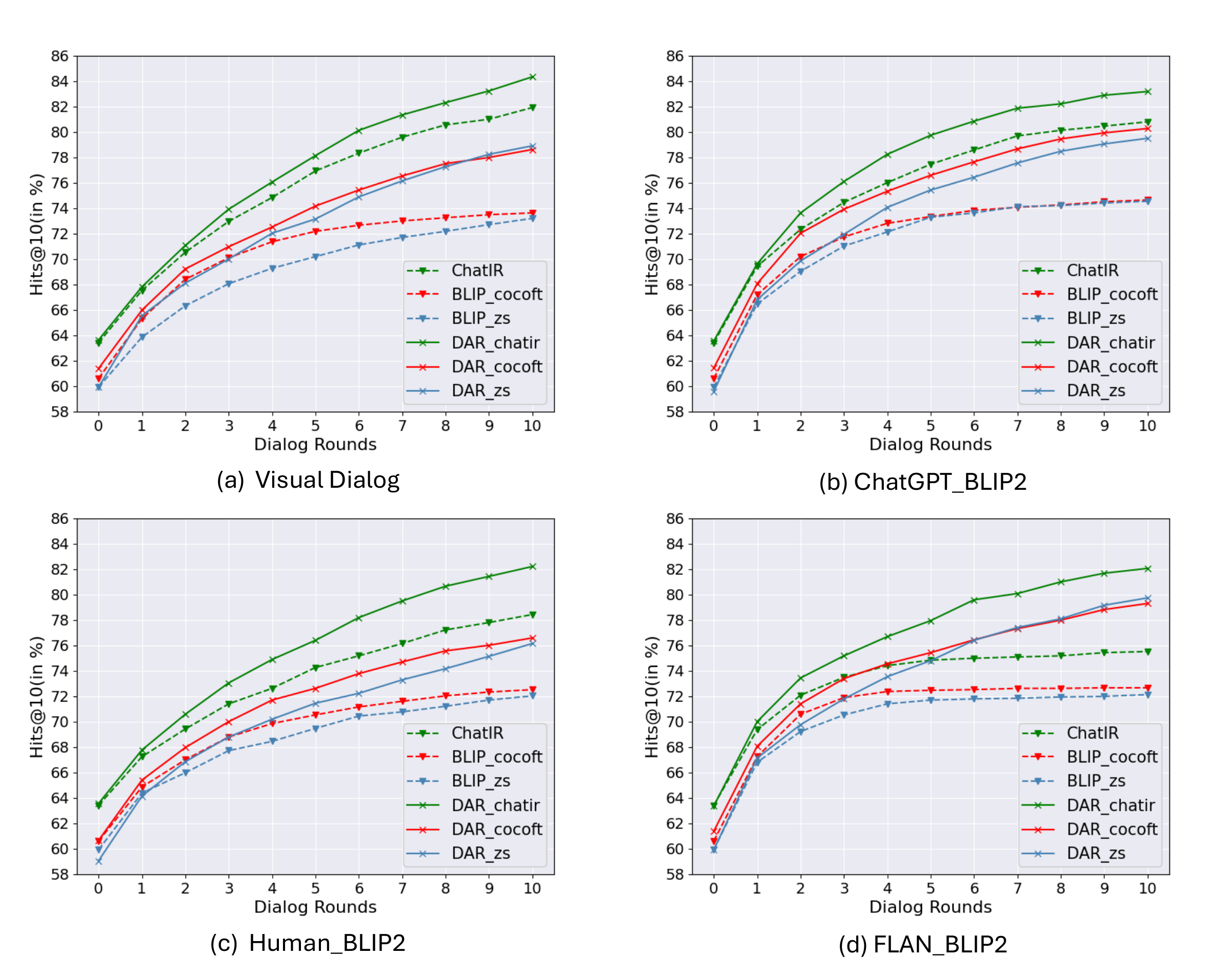}
    \vspace{-3mm}
    \caption{The experimental results for four evaluated benchmarks. Note that for Hits@10, a higher value is better. It is an accumulative metric because we cease to use additional dialogues once the image attains a top-k rank.
reduction in performance.}
    \label{fig:result1}
    \vspace{-3mm}
\end{figure*}


\section{Experimental Results}
\label{sec:exp}

\subsection{Experimental Settings}
\label{sec:exp_setting}

We evaluate our proposed DAR framework in interactive retrieval settings using four well-established benchmarks. Specifically, we employ the validation set of Visual Dialog (VisDial)~\cite{RN625} dataset and three dialog datasets\footnote{Available at: \url{https://github.com/levymsn/ChatIR}} constructed by \cite{RN626}, named \emph{ChatGPT\_BLIP2}, \emph{HUMAN\_BLIP2}, and \emph{Flan-Alpaca-XXL\_BLIP2}. For the latter three dialog datasets, the first part of each name indicates the questioner model used to generate the dataset, and the second part refers to the answer model. Note that \emph{human} refers to professional human assessors. Further details on these three datasets can be found in \cite{RN626}. All four dialog datasets consist of 2{,}064 dialogues, each with 10 dialogue turns.

Following previous work in the interactive cross-modal retrieval domain, we use BLIP~\cite{RN463} as our default MLLM encoder, given its established zero-shot performance and to ensure a fair comparison with prior studies. Unless otherwise specified, we report Hits@10 as our primary evaluation metric.

We adopt the Stable Diffusion~3 model (SD3)~\cite{RN624} as the default diffusion model (DM) in our experiments. Additionally, BLIP-3~\cite{RN627} is employed as the Large Language Model (LLM) for reformulating textual content. Using specially designed prompts, BLIP-3 operates under two distinct reformulation pipelines: one for adapting the dialogue and another for producing aligned prompts for the DM, as described in Section~\ref{sec:reformulation}. We empirically set the weighting factors for textual and visual content to \(0.7\) and \(0.3\), respectively, for the first two dialogue turns. Starting from turn 3, the weights are both set to \(0.5\). This strategy balances the influence of textual vs.\ visual embeddings as the dialogue becomes more dynamic. In all experiments, we fix the number of generated images per turn at three. Code for all experiments is available at \url{https://github.com/longkukuhi/Diffusion-Agumented-Retrieval}.

\textbf{Baselines and DAR variants.} We compare our proposed DAR framework with three baselines, namely ChatIR, ZS, and COCOFT:

\begin{itemize}
    \item ChatIR:
    We adopt the BLIP-based variant of ChatIR~\cite{RN626} as our baseline. Since it is finetuned on the Visual Dialog dataset, ChatIR represents a finetuned model that helps us compare both the effectiveness and efficiency benefits of DAR.

    \item ZS:
    The zero-shot (ZS) baseline uses BLIP with its original, publicly available pretrained weights~\footnote{Available at: \url{https://github.com/salesforce/BLIP}}, without any finetuning on retrieval datasets. This setup captures the common scenario where researchers or practitioners directly rely on publicly released weights without task-specific training. 

    \item COCOFT:
    The COCOFT baseline denotes the BLIP model finetuned on the popular MSCOCO~\cite{RN467} retrieval dataset. Although it benefits from MSCOCO-specific training, it remains non-finetuned for interactive text-to-image retrieval (I-TIR). Consequently, dialogues including questions and answers are applied directly as queries for retrieval. This serves as an indicator of how previously finetuned single-turn retrieval models perform in an interactive retrieval setting.

\end{itemize}

We integrate each of the three baseline encoders into our DAR framework to evaluate improvements across different scenarios, denoting them as \texttt{DAR\_xxx}. For example, \texttt{DAR\_chatir} adopts the BLIP-based ChatIR model as the encoder. Notably, comparisons should primarily focus on the corresponding pairs—such as \texttt{DAR\_chatir} versus ChatIR—whose lines are represented by the same colors in Figure \ref{fig:result1}. \zj{This is because these pairs are initialized from the same pre-trained weights, ensuring a fair comparison that shows how the addition of DAR affects performance.}

\vspace{-2mm}
\subsection{Zero-Shot I-TIR Performance}
We first investigate how our proposed DAR framework performs under zero-shot I-TIR conditions. Specifically, \texttt{BLIP\_zs} denotes the baseline where BLIP is used with its original, publicly available pretrained weights—without any finetuning on retrieval datasets. Our \texttt{DAR\_zs} setup similarly employs pretrained BLIP as the encoder, thereby reflecting a common scenario in which researchers or practitioners rely solely on publicly released weights.

Each subfigure (a–d) in Figures~\ref{fig:result1} presents the performance of \texttt{DAR} variants and baseline models across the four benchmarks outlined in Section~\ref{sec:exp_setting}, evaluated over multiple conversational turns.

In this section, we compare the dashed blue lines (baseline \texttt{BLIP\_zs}) with the solid blue line (\texttt{DAR\_zs}) across the four subfigures (a–d) in Figures~\ref{fig:result1}. We can tell that \texttt{DAR\_zs} consistently outperforms \texttt{BLIP\_zs} across all four evaluated benchmarks. Notably, the FLAN\_BLIP2 dataset exhibits the largest improvement, with \texttt{DAR\_zs} achieving a 7.61\% increase in Hits@10 after 10 dialog rounds. Indeed, the performance gap between \texttt{DAR\_zs} and \texttt{BLIP\_zs} consistently widens as the dialogue extends over multiple turns.

These findings demonstrate that DAR effectively boosts zero-shot performance in interactive retrieval tasks where our ``query" is derived from a complex and diverse dialog, and it can do this without incurring additional tuning overhead.

\vspace{-2mm}
\subsection{Robustness to Complex and Diverse Dialogue Queries}
\label{sec:robustness}
Since our proposed DAR framework is aimed at tackling the challenges of complex and diverse dialogue queries in I-TIR, evaluating its robustness under such conditions is important. \zj{Therefore, we focus our analysis on the most challenging benchmark among the four we evaluated—specifically the one where the best-performing model achieves the lowest Hits@10 score.} In this section, we analyze the solid green lines (\texttt{DAR\_chatir}) as the best-performing model, comparing them with the dashed lines representing the three baselines across the four subfigures (a–d) in Figures~\ref{fig:result1}.

We observe a similar performance trend across three of the benchmarks, whereas FLAN\_BLIP2 differs considerably (see Figure~\ref{fig:result1}.d). On the FLAN\_BLIP2 benchmark, three baseline models (dashed lines) without DAR tend to see their performance plateau around the fifth dialogue turn, indicating they cannot effectively leverage additional complexities introduced later in the conversation. In contrast, our DAR framework continues to improve after turn~5, showing no signs of saturation. Additionally, even our best-performing model, (\texttt{DAR\_chatir}), achieves the lowest Hits@10 on the FLAN\_BLIP2 benchmark. We attribute this to the highly diverse dialogues in FLAN\_BLIP2 compared to the other datasets, making it an important test case for evaluating out-of-distribution performance.

This suggests that DAR can better exploit the extra information presented in later dialogue turns and generate images more closely aligned with the user’s target. Therefore, DAR demonstrates greater robustness to complex and diverse dialogue queries, fulfilling the requirements of I-TIR.

\vspace{-3mm}
\subsection{Gains and losses of DAR compared to finetuned models}
In this section, we investigate both the effectiveness and efficiency of the zero-shot, non-finetuned version of our proposed DAR framework, \texttt{DAR\_zs}, by comparing it against the finetuned \texttt{ChatIR} baseline. Thus, we compare the solid blue line (\texttt{DAR\_zs}) with the dash green line (\texttt{ChatIR}) across the four subfigures (a–d) in Figures~\ref{fig:result1}. Overall, \texttt{DAR\_zs} demonstrates competitive performance on all evaluated benchmarks.

As analyzed in Section~\ref{sec:robustness}, \emph{FLAN\_BLIP2} is the most challenging dataset in our experiments, owing to its particularly complex and dynamic interactive dialogues. Notably, \texttt{DAR\_zs} (solid blue line) outperforms the finetuned \texttt{ChatIR} (dash green line) model by 4.22\% in Hits@10 at turn 10 (Figure~\ref{fig:result1}.d), supporting our hypothesis that finetuning MLLMs on limited I-TIR datasets can undermine their ability to handle out-of-distribution queries. Finetuned models like \texttt{ChatIR} perform best on the VisDial dataset (see Figure~\ref{fig:result1}.a), given their extensive finetuning on 123k VisDial training samples. Even in this favorable setting, the worst-case of non-finetuned DAR model, \texttt{DAR\_zs}, lags behind \texttt{ChatIR} by only 3.01\% in Hits@10, all while saving 100\% of the finetuning time.

This suggests that ChatIR’s narrower finetuned distribution makes it more prone to out-of-distribution errors during inference, leading to underperformance. In contrast, DAR excels in these failure scenarios and remains competitive even on ChatIR’s own finetuned dataset.

\vspace{-4mm}
\subsection{Compatibility with finetuned Models}
Our proposed DAR can also be combined with various encoders, including those already finetuned for dialogue-based text. In this context, the main comparison is between \texttt{DAR\_chatir} and \texttt{ChatIR}, where \texttt{DAR\_chatir} employs the finetuned ChatIR model as its encoder. Thus, we compare the solid green line (\texttt{DAR\_chatir}) with the dash green line (\texttt{ChatIR}) across the four subfigures (a–d) in Figures~\ref{fig:result1}.

Notably, as shown in Figures~\ref{fig:result1}, \texttt{DAR\_chatir} consistently outperforms \texttt{ChatIR} without additional training. Because \texttt{ChatIR} is finetuned on the VisDial dataset, it holds an obvious advantage in that benchmark. Building upon this strong baseline, \texttt{DAR\_chatir} still achieves a 2.42\% improvement in Hits@10 over \texttt{ChatIR} on the VisDial dataset. 

More importantly, as discussed earlier, finetuned models often struggle with out-of-distribution data in interactive retrieval due to the inherently dynamic nature of multi-turn dialogues. This challenge arises in the other three benchmark datasets we evaluate. Consequently, \texttt{DAR\_chatir} exhibits even greater performance gains in these scenarios, achieving up to a 9.4\% increase in Hits@10 compared to \texttt{ChatIR}.

These findings support our claim that DAR can enhance zero-shot retrieval performance without further training—particularly in the face of dynamic dialogues in interactive retrieval—even when the underlying encoder is finetuned on a specific dataset.

\vspace{-2mm}
\subsection{Does finetuning on Single-Turn Retrieval Datasets Improve Performance?}
Given that single-turn and interactive text-to-image retrieval share some similarities at turn 0, this section investigates whether finetuning on a single-turn retrieval dataset—specifically, the widely used MSCOCO dataset \cite{RN467}—enhances performance in interactive retrieval tasks.

As shown in all four subfigures of Figure~\ref{fig:result1}, the benefit of finetuning on a single-turn retrieval dataset is relatively limited. While finetuned models such as \texttt{BLIP\_cocoft} (dash red lines) and \texttt{DAR\_cocoft} (solid red lines) exhibit noticeable performance improvements at earlier turns compared to their zero-shot counterparts (\texttt{BLIP\_zs} and \texttt{DAR\_zs}), this advantage diminishes as the dialogue progresses. By turn 8, the zero-shot model \texttt{DAR\_zs} even surpasses the COCO finetuned model \texttt{DAR\_cocoft} in performance, as illustrated in Figure~\ref{fig:result1}.d.

These findings show that conventional finetuning strategies on single-turn datasets are insufficient to address the increasing query diversity and complexity inherent in interactive text-to-image retrieval. This underscores the limitations of finetuning-based approaches in handling evolving multi-turn interactions. In contrast, our proposed DAR framework effectively bridges this research gap by enhancing zero-shot performance without requiring additional finetuning, making it a more scalable and adaptive solution for real-world retrieval scenarios.

\begin{figure*}[]
    \centering
    \includegraphics[width=0.75\linewidth]{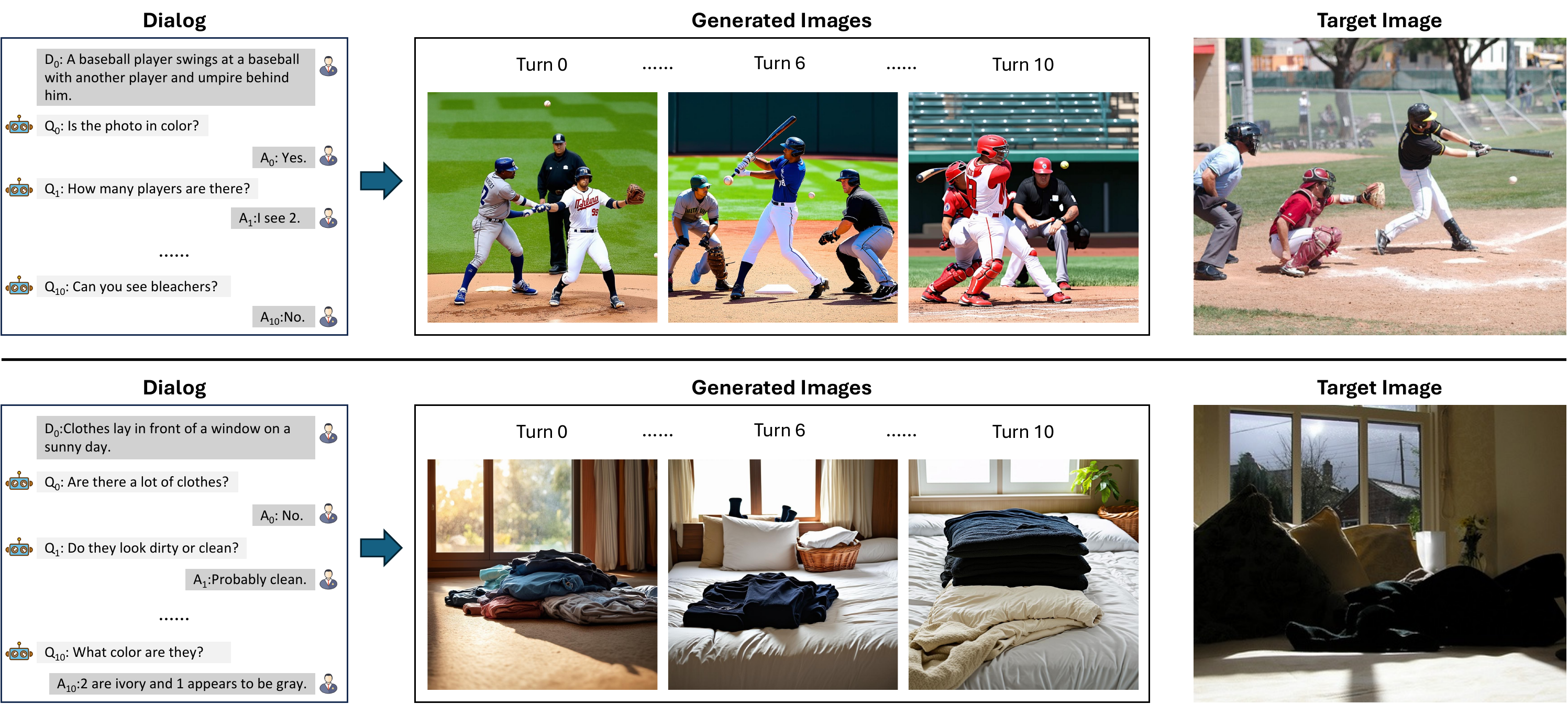}
    \vspace{-3mm}
    \caption{The examples of generated images in DAR.}
    \label{fig:dm_exm}
    \vspace{-4mm}
\end{figure*}

\vspace{-2mm}
\section{Analysis}
\vspace{-0.5mm}
\subsection{Qualitative Analysis of DAR-Generated Images}
\label{sec:gen_example}
Beyond the numerical results presented in Section~\ref{sec:exp}, we conduct an in-depth analysis of the images generated by DAR and uncover several key insights. As illustrated in Figure~\ref{fig:dm_exm}, the initial generated images tend to lack fine details and are easily distinguishable as synthetic. For instance, as the dialogue progresses and additional contextual information is incorporated, the generated images become increasingly photorealistic, with details aligning more closely with the target image—such as the accurate color of the player's clothing. A similar trend is observed in the second-row example, where the diffusion model dynamically adjusts clothing colors based on dialogue updates, ultimately enhancing retrieval accuracy.

These findings highlight DAR’s ability to iteratively refine image generation based on evolving dialogue cues, enabling more precise semantic alignment with the target image. This adaptive generation process not only improves retrieval performance but also demonstrates the potential of integrating generative models into cross-modal retrieval frameworks.

\vspace{-2mm}
\subsection{Compatibility of DAR with Different Encoders and Generators}
\label{sec:compability}
To assess the compatibility of DAR with different MLLMs as encoders and DMs as visual generators, we evaluate its performance using two widely adopted MLLMs—CLIP \cite{RN471} and BEiT-3 \cite{RN289}—as encoders, as well as Stable Diffusion v2-1 \cite{Rombach_2022_CVPR} as an alternative visual generator.

The CLIP model, constrained by its maximum input length of 77 tokens, struggles to handle complex and lengthy dialogue-based queries compared to simpler caption-based queries in single-turn cross-modal retrieval tasks. Consequently, CLIP achieves a peak performance of \textbf{56.64\% Hits@10} on the Visual Dialog benchmark. Despite this limitation, incorporating CLIP as the encoder within DAR still yields an improvement of 5.31\% Hits@10 after 10 dialogue turns, demonstrating the robustness of our framework even with models that have constrained input capacity.

BEiT-3, a stronger encoder than CLIP and BLIP models, further enhances performance. When employed as the encoder within DAR, the framework achieves a \textbf{6.28\% improvement in Hits@10} after 10 turns on the Visual Dialog benchmark, compared to the standalone pre-trained BEiT-3 model, highlighting the adaptability of DAR in leveraging stronger backbone encoders.

In addition to evaluating different encoders, we also assess the impact of using Stable Diffusion v2-1 as the visual generator. While DAR continues to yield notable improvements, the performance gain is slightly lower compared to using Stable Diffusion 3 (SD3), aligning with SD3’s superior generative capabilities. Specifically, DAR achieves a 6.37\% Hits@10 with Stable Diffusion v2-1, compared to 7.61\% Hits@10 with SD3.

These results demonstrate that DAR is flexible and compatible with various encoder and generation models. Regardless of the choice of encoder or visual generator, DAR consistently enhances retrieval performance, reinforcing its generalizability as an effective framework for zero-shot cross-modal retrieval.

\vspace{-3.5mm}
\subsection{Impact of the Number of Generated Images on Performance}
In our evaluation, even when generating only a single image within our framework, we observe a \textit{6.43\%} improvement in Hits@10 on the \texttt{FLAN\_BLIP2} benchmark compared to the ChatIR baseline. When increasing the number of generated images to three, performance further improves to \textit{7.61\%}. However, beyond this point, we observe diminishing returns, with Hits@10 reaching saturation as the number of generated images continues to increase. 

Considering the trade-off between effectiveness and computational efficiency, we set three generated images as the default configuration in DAR to achieve an balance between performance and inference cost.

\vspace{-3mm}
\subsection{Generation Overhead}
While DAR eliminates the need for expensive finetuning, it introduces additional inference-time overhead due to the reformulation and generation of visual content. However, this trade-off is small compared with the benefits in retrieval performance and adaptability. In our experiments using a single Nvidia RTX 4090 GPU, the image generation process takes approximately \textit{5 seconds}, while the query reformulation step incurs only \textit{0.5 seconds} of additional processing time.

\zj{To put this into perspective, ChatGPT-\texttt{o1} and other `reasoning' LLMs use additional reasoning steps, which enhances response quality at the cost of increased inference time (often exceeding \textit{10 seconds}). Analogously, DAR achieves substantial gains in zero-shot retrieval while entirely eliminating training costs, making it an efficient and scalable alternative. Consequently, the modest inference overhead is a worthwhile trade-off, particularly in applications where adaptability and retrieval quality are crucial.}

\vspace{-2mm}
\section{Conclusion}
In this work, we introduce DAR, a novel framework that eliminates the need for finetuning multimodal large language models (MLLMs) for Interactive Text-to-Image Retrieval (I-TIR), thereby preserving their generalizability. Extensive experiments show that DAR performs competitively with existing I-TIR models, whether they are fine-tuned or pretrained. Our analysis reveals that finetuning MLLMs on limited retrieval datasets compromises their ability to handle out-of-distribution queries. Furthermore, our results illustrate that generating multiple intermediate, multi-faceted representations of user intent enables a many-to-one mapping between text and images, effectively accommodating the diverse and complex queries characteristic of I-TIR. This work highlights the potential of diffusion-augmented retrieval and suggests avenues for future exploration in optimizing efficiency, supporting multimodal queries, and extending to real-world applications.


\clearpage
\bibliographystyle{ACM-Reference-Format}
\bibliography{sample-base}

\appendix

\end{document}